\documentclass[aps,preprint,floatfix]{revtex4-1}
\usepackage{graphicx}
\usepackage{amsmath}
\usepackage{makeidx}
\usepackage{amsfonts}
\usepackage{amssymb}

\begin{document}

\title{Multilayer wave functions: A recursive coupling of local excitations}

\author{A. Ramezanpour}
\affiliation{Department of Applied Science and Technology, Politecnico di Torino,
Corso Duca degli Abruzzi 24, 10129 Torino, Italy}

\email{aramezanpour@gmail.com}

\date{\today}

\begin{abstract}
Finding a succinct representation to describe the ground state of a disordered interacting system could be very helpful in understanding the interplay between the interactions that is manifested in a quantum phase transition. In this work we use some elementary states to construct recursively an ansatz of multilayer wave functions, where in each step the higher-level wave function is represented by a superposition of the locally "excited states" obtained from the lower-level wave function. This allows us to write the Hamiltonian expectation in terms of some local functions of the variational parameters, and employ an efficient message-passing algorithm to find the optimal parameters. We obtain good estimations of the ground-state energy and the phase transition point for the transverse Ising model with a few layers of mean-field and symmetric tree states. The work is the first step towards the application of local and distributed message-passing algorithms in the study of structured variational problems in finite dimensions.        
\end{abstract}


\maketitle

\section{Introduction}\label{S0}
In a wave function approach to the study of an interacting quantum system we usually resort to some physical and numerical insights to suggest a reasonable variational wave function that approximates the quantum state of the system. In fact, providing a succinct representation that well describes the physical state of the system means we know how to model the relevant quantum correlations in an efficient way. The number of variational parameters we need to characterize such a wave function could be of the order of the size of system, depending on the nature of quantum correlations captured by the wave function. Here it is essential to have an efficient optimization algorithm for minimizing the Hamiltonian expectation over the space of the variational parameters.     

In this work we will use some ideas from the physics of quantum many-body systems, more specifically the matrix product states~\cite{FNW-cmp-1992,OR-prl-1995} and the coupled cluster method~\cite{C-np-1958,CK-np-1960}, to construct an ansatz of multilayer wave functions for a possibly disordered quantum system of interacting spins. The matrix product states and the related generalizations~\cite{SDV-pra-2006,VMC-advp-2008}, e.g. multiscale entanglement-renormalization ~\cite{V-prl-2007,V-prl-2008} and projected entangled pair states~\cite{VC-2004,MVC-pra-2007}, can be constructed by integrating over some auxiliary degrees of freedom interconnected in a specific manner to the physical variables to account for the entanglement in different parts of the system~\cite{H-prb-2006,H-prb-2007}. On the other hand, in the coupled cluster method one starts from an appropriate reference state, e.g. the Hartree-Fock wave function, and elaborates on the local excitations to obtain more accurate wave functions and estimations for the ground-state energy~\cite{B-ln-1998}. 
  
It is always useful in the study of interacting systems to start with the mean-field (MF) wave functions (or product states). More accurate wave functions are obtained by adding interactions between the variables~\cite{J-pr-1955,HE-prl-1988,IOD-prb-2009,CKUC-prb-2009}. Here, in general, we have to resort to some approximation algorithms, e.g. Monte Carlo~\cite{MC-rmp-2001}, to compute efficiently the quantum expectations. In Ref.~\cite{R-prb-2012} we proposed to estimate the expectations within the Bethe approximation, which allows us to write the Hamiltonian expectation in terms of some local functions of the variational parameters and the cavity marginals of the Bethe approximation; for a review of similar methods see~\cite{BFKSZ-pr-2013}. Note that the Bethe estimation of the Hamiltonian expectation is not necessarily an upper bound for the ground-state energy, unless the interaction graph defined by the trial wave function has a tree structure. Nevertheless, the same approximation offers efficient message-passing algorithms that have been proved useful in the study of random constraint satisfaction and optimization problems~\cite{MPZ-science-2002,MZ-pre-2002,MM-book-2009}. 
  
Symmetric wave functions with a tree structure provide us with another category of computationally tractable states which somehow complement the mean-field states; while the latter wave functions are good candidates for the state of the system in the ordered (ferromagnetic, or localized) phase, the symmetric states are more appropriate in the disordered (paramagnetic, or extended) phase. Moreover, in both the cases we can easily construct an orthonormal set of locally excited states that could be useful in the framework of the coupled cluster method~\cite{BR-jstat-2013}. We remark that the weighted graph states studied in quantum physics and information theory can be represented by application of some two-body unitary operators on initially mean-field states~\cite{HDERNB-2006}. Similarly, we can obtain a weighted graph state with an initially tree wave function and still compute efficiently the quantum expectation of local observables. 

In this study we use the mean-field and the symmetric tree wave functions to construct an ansatz of multilayer wave functions by a recursive coupling of the local excitations; we start from a reference wave function and in each step we construct a higher-level wave function by taking a superposition of the locally "excited states" obtained from the wave function in the previous step. In the last step we minimize the Hamiltonian expectation with respect to the variational parameters characterizing the reference state and the superposition functions. Note that we use the "excited state" for any state that is orthogonal to the trail wave function; in this sense, the average energy of an excited state could be less than that of the reference wave function as we minimize the energy over the whole set of the parameters only in the end of the process. The simple structure of the wave functions allows us to work with local energy functions of the variational parameters which is essential for utilizing distributed message-passing algorithms in the study of the optimization problem. In principle, the method can be implemented with more general wave functions for spin systems with an arbitrary interaction graph.

In the following we will specify the wave functions and the local excitations that we are going to work with in the multilayer wave functions. Then we present the message-passing algorithm that is used to minimize the Hamiltonian expectation, and report some preliminary results for the ferromagnetic transverse Ising model in one and two spatial dimensions.

\section{Definitions}\label{S1}
Consider the transverse Ising model with Hamiltonian  
\begin{align}
H =-\sum_{(ij) \in \mathcal{E}_q} J_{ij} \sigma_i^z\sigma_j^z-\sum_{i} h_i \sigma_i^x,
\end{align} 
where $i=1,\dots,N$ labels the sites in the quantum interaction graph $\mathcal{E}_q$. The $\sigma_i^{x,y,z}$ are the standard Pauli matrices. 
And we use the orthonormal set of states $|\boldsymbol\sigma \rangle=|\sigma_1,\dots,\sigma_N \rangle$ with $\sigma_i=\pm 1$ in the $\sigma_i^{z}$ representation. 

Starting from a reference wave function $| \Psi_0 \rangle= \sum_{\boldsymbol\sigma} \psi_0(\boldsymbol\sigma;\mathbf{P}^0) |\boldsymbol\sigma \rangle$ characterized by the variational parameters $\mathbf{P}^0$, we construct the orthonormal set of excited states $\mathcal{S}_0\equiv \{| \Psi_{0,s_0} \rangle | s_0=0,\dots,\mathcal{N}_0\}$, where $| \Psi_{0,0} \rangle  \equiv | \Psi_0 \rangle$. Then a higher-level wave function is obtained by taking a superposition of the excited states $| \Psi_1 \rangle = \sum_{s_0} \psi_1(s_0;\mathbf{P}^1) | \Psi_{0,s_0} \rangle$. Note that $| \Psi_1 \rangle$ depends also on $\mathbf{P}^0$ through the $| \Psi_{0,s_0} \rangle$. The process can be repeated for $t$ steps to construct a $(t+1)$-layer wave function. At layer $t$ we have $| \Psi_t \rangle = \sum_{s_{t-1}} \psi_t(s_{t-1};\mathbf{P}^t) | \Psi_{t-1,s_{t-1}} \rangle$ with the orthonormal set of excited states $\mathcal{S}_{t-1}\equiv \{| \Psi_{t-1,s_{t-1}} \rangle | s_{t-1}=0,\dots,\mathcal{N}_{t-1}\}$ and $| \Psi_{t-1,0} \rangle \equiv  | \Psi_{t-1} \rangle$. The aim is to minimize the Hamiltonian expectation over the variational parameters,
\begin{align}
E_0= \min_{\{\mathbf{P}^0,\dots,\mathbf{P}^t\}}  \langle \Psi_t | H | \Psi_t \rangle,
\end{align}
for some succinct representation of the variational states characterized by the parameters and the nature of excitations in the excited states. We recall that by the "excited state" we mean any state that is orthogonal to the trial wave function, and that is not necessarily an eigenstate of the Hamiltonian.

One can write the excited states at layer $l>0$ as $| \Psi_{l,s_{l}} \rangle=\sum_{s_{l-1}} \psi_{l}(s_{l-1},s_{l};\mathbf{P}^{l})| \Psi_{l-1,s_{l-1}} \rangle$. Notice that $\psi_{l}(s_{l-1};\mathbf{P}^{l})=\psi_{l}(s_{l-1},0;\mathbf{P}^{l})$ as defined above. Moreover, by the orthogonality of the excited states we have $\sum_{s_{l-1}}\psi_{l}^*(s_{l-1},s_l';\mathbf{P}^{l})\psi_{l}(s_{l-1},s_l;\mathbf{P}^{l})=\delta_{s_l',s_l}$. This results to the following wave function at layer $t$:
\begin{align}
| \Psi_t \rangle= \sum_{s_{t-1},s_{t-2},\dots,s_0,\boldsymbol\sigma} \psi_{t}(s_{t-1};\mathbf{P}^{t})\psi_{t-1}(s_{t-2},s_{t-1};\mathbf{P}^{t-1}) \dots \psi_{1}(s_{0},s_{1};\mathbf{P}^{1})\psi_{0}(\boldsymbol\sigma,s_{0};\mathbf{P}^{0})| \boldsymbol\sigma \rangle.
\end{align}
Then the average value of a local operator $O$ with matrix elements $O^{\boldsymbol\sigma'\boldsymbol\sigma} \equiv \langle \boldsymbol\sigma '|O| \boldsymbol\sigma \rangle$ can be computed in a recursive way by,
\begin{align}
\langle O \rangle_t = \sum_{s_{t-1}} |\psi_{t}(s_{t-1};\mathbf{P}^{t})|^2 
\left( \sum_{s_{t-1}'} \frac{\psi_{t}^*(s_{t-1}';\mathbf{P}^{t})}{\psi_{t}^*(s_{t-1};\mathbf{P}^{t})} [ O ]_{t-1}^{s_{t-1}'s_{t-1}} \right),
\end{align}
where
\begin{align}\label{matrix}
[ O ]_l^{s_l's_l} \equiv  \sum_{s_{l-1}} |\psi_{l}(s_{l-1},s_l;\mathbf{P}^{l})|^2 
\left( \sum_{s_{l-1}'} \frac{\psi_{l}^*(s_{l-1}',s_l';\mathbf{P}^{l})}{\psi_{l}^*(s_{l-1},s_l;\mathbf{P}^{l})} [ O ]_{l-1}^{s_{l-1}'s_{l-1}} \right),
\end{align}
setting $s_{-1} \equiv \boldsymbol \sigma$ and $[ O ]_{-1} \equiv O$.
Obviously, to compute the Hamiltonian expectation efficiently we have to limit ourselves to simple enough wave functions and excitations.

\section{characterizing the wave functions}\label{S2}
A trial wave function $|\Psi \rangle=\sum_{\boldsymbol\sigma} \psi(\boldsymbol\sigma;\mathbf{P}) |\boldsymbol\sigma \rangle$ is characterized by the structure of the coefficients and the set of parameters $\mathbf{P}$. A correlated wave function can be constructed by considering the one-body and the two-body interactions~\cite{J-pr-1955,HE-prl-1988,IOD-prb-2009,CKUC-prb-2009},
\begin{align}
\psi(\boldsymbol\sigma;\mathbf{P}) \propto \prod_{i} \phi_{i}(\sigma_i;P_i) \prod_{(ij)\in \mathcal{E}} \phi_{ij}(\sigma_i,\sigma_j;P_{ij}).
\end{align} 
In appendix \ref{app-CW} we describe an approximation algorithm to estimate the quantum expectations for such correlated wave functions.  

In the study of multilayer wave functions we will work with the mean-field and the symmetric tree states.  The mean-field states can in general be represented by
\begin{align}\label{MFS}
\psi(\boldsymbol\sigma;\mathbf{B})=e^{\hat{i}\Theta(\boldsymbol\sigma)}\prod_i \left(\frac{e^{B_i\sigma_i/2}}{\sqrt{2\cosh(B_i^R)}} \right),
\end{align}
with an arbitrary real phase $\Theta(\boldsymbol\sigma)$ and complex fields $B_i=B_i^R+\hat{i}B_i^I$. We call such a state mean-field because the probability measure $\mu(\boldsymbol\sigma;\mathbf{B})\equiv |\psi(\boldsymbol\sigma;\mathbf{B})|^2$ represents a classical system of independent variables.
On the other hand we have the symmetric tree states:
\begin{align}\label{SS}
\psi(\boldsymbol\sigma;\mathbf{K})=\frac{e^{\hat{i}\Theta(\boldsymbol\sigma)}}{\sqrt{2^N}}\prod_{(ij) \in \mathcal{T}} \left(\frac{e^{K_{ij}\sigma_i\sigma_j/2}}{\sqrt{\cosh(K_{ij}^R)}} \right),
\end{align}
for some tree interaction graph $\mathcal{T}$ and complex couplings $K_{ij}=K_{ij}^R+\hat{i}K_{ij}^I$. Here the associated probability measure  has a tree structure and is symmetric under $\boldsymbol \sigma \to -\boldsymbol \sigma$. 
In both the cases we will write the phase in terms of some local interactions: $\Theta(\boldsymbol\sigma)=\sum_i \Lambda_i \sigma_i/2+\sum_{(ij) \in \mathcal{E}_I} \Gamma_{ij} \sigma_i\sigma_j/2$. Note that besides the variational parameters we need to specify the interaction graphs $\mathcal{E}_I$ and $\mathcal{T}$. Here to maximize the gain from the interactions in the wave functions we follow the quantum interaction graph $\mathcal{E}_q$; i.e., we prefer to have interactions between the nearest neighbors in $\mathcal{E}_q$, and then between the next nearest neighbors and so on.

\section{Characterizing the excitations}\label{S3}
In this section we will take the mean-field and the symmetric tree states
to illustrate the nature of the local excitations that we are going to exploit in constructing the multilayer wave functions.
  
\subsection{Local excitations in the mean-field states}\label{S31}
Consider the mean-field state $|\Psi \rangle= \sum_{\boldsymbol\sigma}\psi(\boldsymbol\sigma)|\boldsymbol\sigma \rangle$ with $\psi(\boldsymbol\sigma)= e^{\hat{i}\Theta(\boldsymbol\sigma)}\prod_i \left(\frac{e^{B_i\sigma_i/2}}{\sqrt{2\cosh(B_i^R)}} \right)$.
We define a set of orthonormal mean-field states $|i_1\dots i_n \rangle$ for $n=1,\dots,N$ with the same phase $\Theta(\boldsymbol\sigma)$ but different fields $\tilde{B}_i$. 
The state $|i_1\dots i_n \rangle$ is orthogonal to $|\Psi\rangle$ at sites $\{i_1\dots i_n\}$; that is $\tilde{B}_i^R=-B_i^{R}$ and $\tilde{B}_i^I=B_i^{I}+\pi$ for $i \in \{i_1\dots i_n\}$, otherwise $\tilde{B}_i=B_i$. We can represent all the above states in the occupation number representation by $|\mathbf{s} \rangle \equiv |s_1,\dots,s_N \rangle$, with $s_i \in \{0,1\}$ to show the presence of a local excitation at site $i$. Note that the state $|\mathbf{0} \rangle$ gives the original mean-field state $|\Psi \rangle$, and $\langle \boldsymbol\sigma |\mathbf{s} \rangle=\psi(\boldsymbol\sigma,\mathbf{s};\mathbf{P})$ as defined in Sec. \ref{S1}. A higher-level wave function is obtained by a superposition of the locally excited states. In appendix \ref{app-mf} we see how the average energy can be computed for such a wave function in the transverse Ising model.

\subsection{Local excitations in the symmetric tree states}\label{S32}
Consider the symmetric tree state $|\Psi \rangle=\sum_{\boldsymbol\sigma} \frac{e^{\hat{i}\Theta(\boldsymbol\sigma)}}{\sqrt{2^N}}\prod_{(ij) \in \mathcal{T}} \left(\frac{e^{K_{ij}\sigma_i\sigma_j/2}}{\sqrt{\cosh(K_{ij}^R)}} \right) |\boldsymbol\sigma \rangle$. We define a set of orthonormal symmetric tree states $|(ij)_1\dots (ij)_n \rangle$ for $n=1,\dots,N-1$ with the same phase $\Theta(\boldsymbol\sigma)$ but different couplings $\tilde{K}_{ij}$. 
The state $|(ij)_1\dots (ij)_n \rangle$ is orthogonal to $|\Psi\rangle$ at edges $\{(ij)_1\dots (ij)_n\}$; that is $\tilde{K}_{ij}^R=-K_{ij}^{R}$ and $\tilde{K}_{ij}^I=K_{ij}^{I}+\pi$ for $(ij) \in \{(ij)_1\dots (ij)_n\}$, otherwise $\tilde{K}_{ij}=K_{ij}$. Again, we represent all the above states in the occupation number representation by $|\mathbf{s} \rangle$, with $s_{ij} \in \{0,1\}$ to show the presence of a local excitation at edge $(ij)$. The state $|\mathbf{0} \rangle$ represents the original tree state $|\Psi \rangle$. In appendix \ref{app-s} we write the Hamiltonian expectation for a superposition of such locally excited states.

Note that instead of having excitations on the edges we could have the excitations on the nodes; here a local excitation on node $i$ is defined by modifying the parameters on all the edges emanating from the node and is represented by the occupation number $s_i$. But, since the number of edges is one less than the number of nodes, we have to use only $N-1$ node variables to represent the orthogonal set of locally excited states.

\section{Optimization algorithm}\label{S4}
In this section we briefly describe a message-passing algorithm to minimize the Hamiltonian expectation and find an estimation of the optimal variational parameters. Let us assume we write the energy as $\langle H \rangle=\sum_a E_a (P_{\partial a})$ where $P_{\partial a}$ is the set of variational parameters $P_v$ that appear in the local energy function $E_a$. To say something about the optimal parameters we can study the following statistical physics problem $\mathcal{Z}=\sum_{\mathbf{P}} e^{-\beta_{opt} \sum_{a}E_a}$. For finite $\beta_{opt}$, we use the Bethe approximation to write the cavity marginals $M_{a\to v}(P_v)$ and $M_{v\to a}(P_v)$ of the variational parameter $P_v$. 
The former messages are sent from the local energy functions to the parameters and give the probability of having the parameter $P_v$ in the absence of the other energy functions involving the parameter,
\begin{align}
M_{a\to v}(P_v) \propto \sum_{\{P_u| u \in \partial a \setminus v\}} e^{ -\beta_{opt}E_a(P_{\partial a})} \prod_{u \in \partial a \setminus v} M_{u\to a}(P_u),
\end{align} 
The latter messages are sent from the parameters to the local energy functions. These messages give the probability of having the parameter $P_v$ in the absence of the $E_a$,
\begin{align}
M_{v\to a}(P_v) \propto \prod_{b \in \partial v \setminus a} M_{b\to v}(P_v).
\end{align} 
Here $\partial v$ is the set of local energy functions depending on $P_v$. 
The above equations are called belief propagation (BP) equations~\cite{KFL-inform-2001,MM-book-2009}.

But we are interested in the limit $\beta_{opt} \to \infty$ where the probability distribution of the variational parameters is concentrated on the optimal ones. Taking the scaling  $M_{a \to v}(P_v) = e^{-\beta_{opt}\mathcal{M}_{a \to v}(P_v)}$ and similarly for $M_{a \to v}(P_v)$, we get   
\begin{align}
\mathcal{M}_{a\to v}(P_v) &= \min_{\{P_u| u \in \partial a \setminus v\}} \left\{ E_a(P_{\partial a}) + \sum_{u \in \partial a \setminus v} \mathcal{M}_{u\to a}(P_u) \right\},\\
\mathcal{M}_{v\to a}(P_v) &= \sum_{b \in \partial v \setminus a} \mathcal{M}_{b\to v}(P_v).
\end{align} 
These are the so called minsum equations~\cite{KFL-inform-2001}. The equations can be solved by iteration starting from random initial messages and updating the messages according to the above equations. After each update we shift the messages by a constant to keep $\min_{P_v}\mathcal{M}_{a \to v}(P_v)=\min_{P_v}\mathcal{M}_{v \to a}(P_v)=0$. Finally, an estimation of the optimal parameters is obtained by  $P_v^{min} = \arg \min  \sum_{a \in \partial v} \mathcal{M}_{a\to v}(P_v)$.

\begin{figure}
\includegraphics[width=10cm]{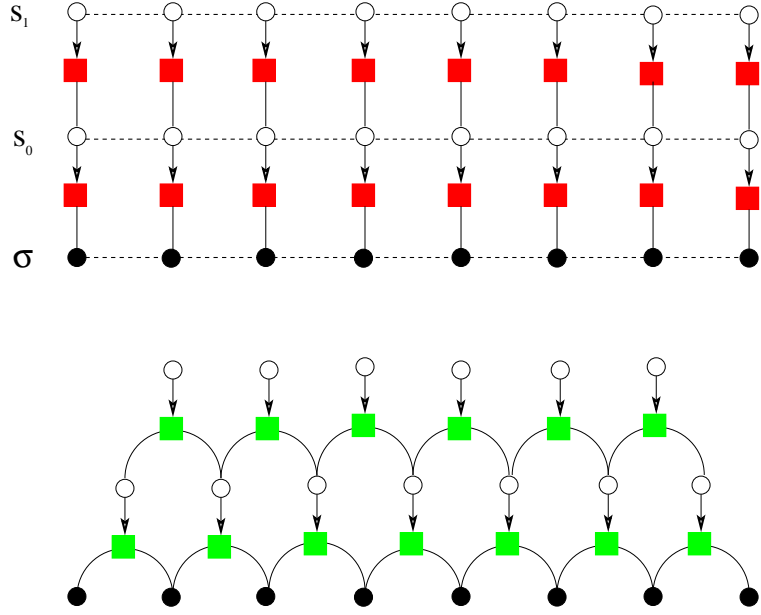}
\caption{ Multilayer wave functions of the mean-field and the symmetric states in one spatial dimension. The filled circles show the physical variables $\boldsymbol\sigma$ and the open circles show the auxiliary variables $\mathbf{s}$. The auxiliary variables $\mathbf{s}_l$ determine the configuration of the local excitations (or interactions) in layer $l-1$. The dashed lines represent the (imaginary) interactions in the phase $\Theta$.}\label{f1}
\end{figure}

\section{Multilayer wave functions of mean-field states}\label{S5}
\subsection{The one-dimensional model}\label{S51}
We start from a mean-field wave function in the one-dimensional system 
and couple the local excitations by another mean-field state as shown in Fig.\ref{f1},
\begin{align}
\psi_0(\boldsymbol \sigma; \mathbf{B}^{0}) &\propto e^{\hat{i}\Theta_0(\boldsymbol\sigma)+\sum_iB_i^0\sigma_i/2},\\
\psi_1(\mathbf{s}_0; \mathbf{B}^{1}) &\propto e^{\hat{i}\Theta_1(\mathbf{s}_0)+\sum_i B_i^1(2s_{0,i}-1)/2}.
\end{align}
For the phases we assume $\Theta_{0}(\boldsymbol\sigma)= \sum_{i} \Gamma_{i,i+1}^{0}\sigma_i\sigma_{i+1}/2$ with some interactions along the quantum interaction graph, and similarly for $\Theta_{1}(\mathbf{s}_0)$. We can still compute exactly the average energies $\langle e_i \rangle=-h_i\langle \sigma_i^x \rangle$ and $\langle e_{i,i+1} \rangle=-J_{i,i+1}\langle \sigma_i^z\sigma_{i+1}^z \rangle$, depending locally on the subset of the parameters 
\begin{align}
\{B_{i-1}^0,\Gamma_{i-1,i}^{0},B_{i}^0,\Gamma_{i,i+1}^{0},B_{i+1}^0;B_{i-2}^1,\Gamma_{i-2,i-1}^{1},\dots,\Gamma_{i+1,i+2}^{1},B_{i+2}^1\}.
\end{align}
Notice that by these average energies we couple the neighboring parameters in different layers. This defines a bipartite interaction graph $\mathcal{E}_v$, where each local energy function $\langle e_i \rangle, \langle e_{i,i+1} \rangle$, represented by node $a$, depends on the parameters $P_{\partial a}$ in its neighborhood subset $\partial a$, and each parameter $P_v \in \{B_i^0,B_i^1,\Gamma_{i,i+1}^0,\Gamma_{i,i+1}^1|i=1,\dots,N\}$, represented by node $v$, appears in the subset of interactions $\partial v$. Given the local energy functions and the dependency graph of the variational parameters we can use the minsum algorithm to minimize the Hamiltonian expectation.

\begin{figure}
\includegraphics[width=10cm]{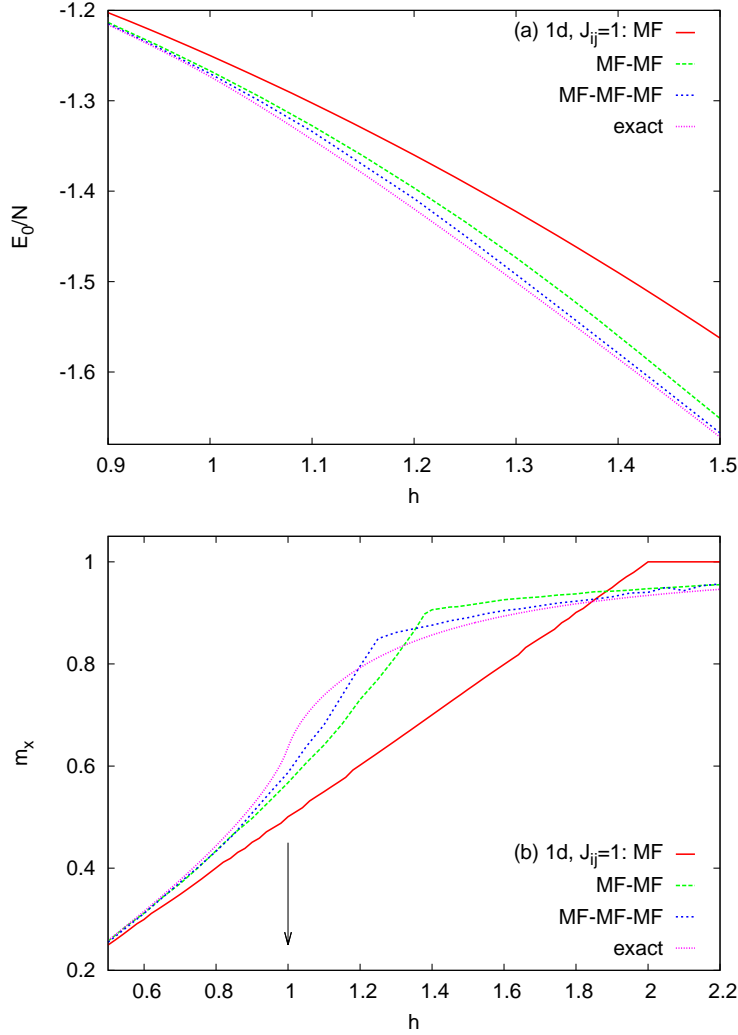}
\caption{ (a) The ground-state energy density $E_0/N$, and (b) the magnetization density $m_x$ for the ferromagnetic transverse Ising model on the infinite one-dimensional lattice obtained by a translationally invariant multilayer wave function of mean-field states. The number of MF refers to the number of layers in the wave function. The arrow shows the exact phase transition point.}\label{f2}
\end{figure}

In the same way, we can continue by coupling the local excitations in layer $t=1$ by another mean-field wave function. In Fig. \ref{f2} we display the results obtained with a few layers of mean-field states for the transverse Ising model in one dimension. The relative error in the ground-state energy $\delta e(h) \equiv (E_0/E_0^{exact}-1)$ computed at the critical point $h=1$ reads $\delta e_{MF}(1)=0.01825(t=0), 0.00499(t=1), 0.00262(t=2)$ for the $(t+1)$-layer wave functions. In the disordered phase for $h=1.1$ we obtain $\delta e_{MF}(1.1)=0.03005(t=0), 0.01143(t=1), 0.00663(t=2)$.

\begin{figure}
\includegraphics[width=10cm]{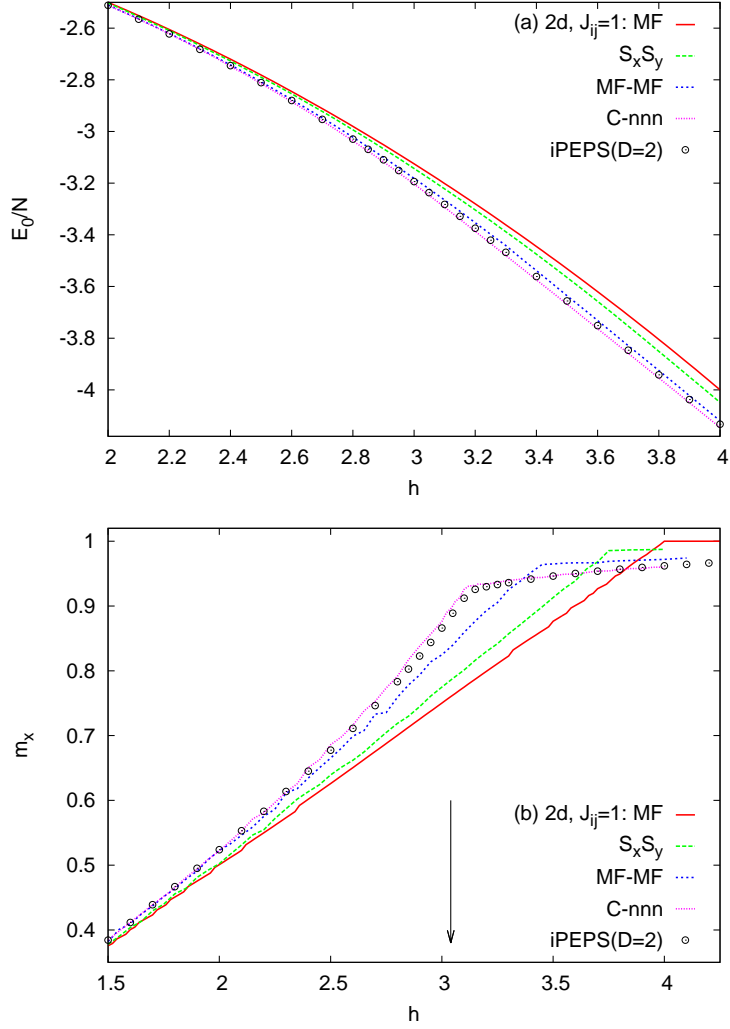}
\caption{(a) The ground-state energy density $E_0/N$, and (b) the magnetization density $m_x$ for the ferromagnetic transverse Ising model on the infinite two-dimensional square lattice obtained by a translationally invariant multilayer wave function of mean-field states and a symmetric tree state. The number of MF refers to the number of layers in the wave function. Here $S_xS_y$ denotes the symmetric tree state with the two sets of $x$ and $y$ auxiliary variables. C-nnn denotes a correlated wave function with nearest and next nearest neighbor interactions along the quantum interaction graph. The open circles are from the iPEPS (infinite projected entangled pair states) algorithm~\cite{JOVVC-prl-2008}. The arrow shows the expected phase transition point from series expansion~\cite{OHZ-jpa-1991}.}\label{f3}
\end{figure}

\subsection{Higher dimensions}\label{S52}
It is straightforward to work with the multilayer wave functions of mean-field states in higher spatial dimensions.  Obviously, the number of variational parameters involved in the quantum expectation of a local observable is of order $(t+1)^{d+1}$ in $d$ dimensions. Consequently, the computation time grows exponentially with $(t+1)^{d+1}$. Note that this complexity is due to the (imaginary) interactions in the phase $\Theta$; in fact if the interaction graph defined by the $\Gamma_{ij}$ is a tree we can compute the quantum expectation of any product operator in a time of order $NC_{max}[2^{2(t+1)}]^2$. Here $C_{max}$ is the maximum connectivity of the nodes in the tree, and $2(t+1)$ is the length of binary string $(\sigma_is_{0,i} \ldots s_{t-1,i};\sigma_i's_{0,i}' \ldots s_{t-1,i}')$ located at each site of the interaction graph. Figure \ref{f3} shows the results we obtain by a two-layer wave function of mean-field states in the two-dimensional transverse Ising model.

\section{Multilayer wave functions of symmetric tree states}\label{S6}
\subsection{The one-dimensional model}\label{S61}
\begin{figure}
\includegraphics[width=10cm]{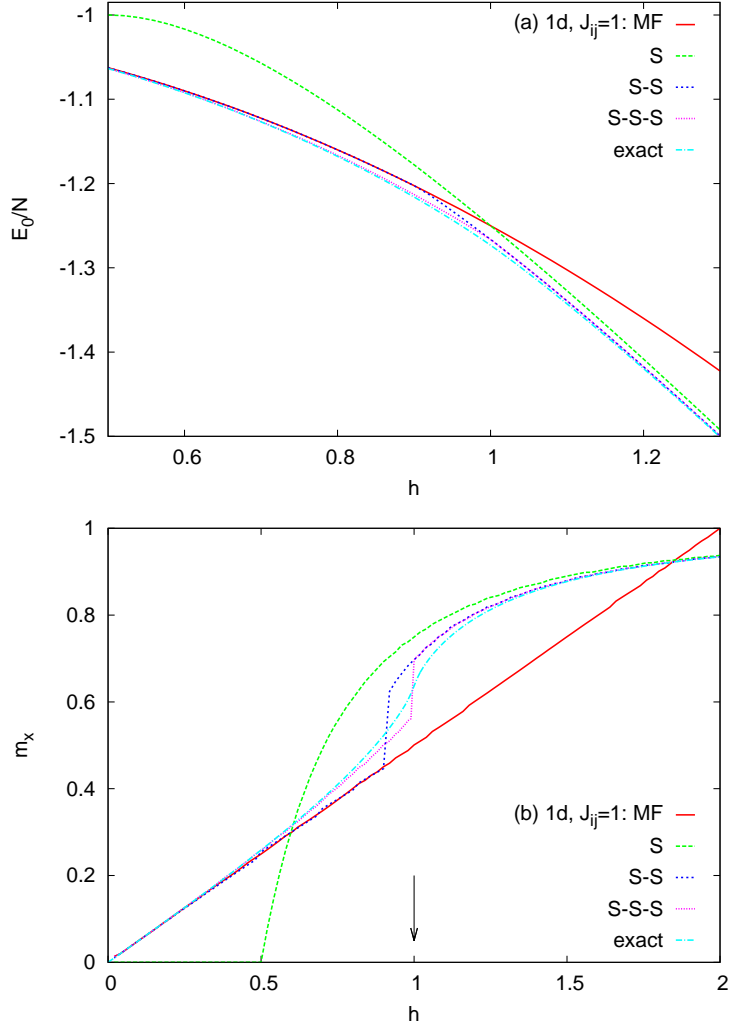}
\caption{ (a) The ground-state energy density $E_0/N$, and (b) the magnetization density $m_x$ for the ferromagnetic transverse Ising model on the infinite one-dimensional lattice obtained by a translationally invariant multilayer wave function of symmetric states. The number of S refers to the number of layers in the wave function. The arrow shows the exact phase transition point. Note that the mean-field states (MF) work better than the symmetric states (S) in the ordered phase, and the symmetric states result to smaller energies  in the disordered phase.}\label{f4}
\end{figure}

We consider the symmetric states in the one-dimensional system as shown in Fig.~\ref{f1}. We take the symmetric wave function $\psi_0(\boldsymbol \sigma; \mathbf{K}^0)\propto e^{\hat{i}\Theta_0(\boldsymbol \sigma)+\sum_{i}K_{i,i+1}^0\sigma_i\sigma_{i+1}/2}$ and couple the local excitations by another symmetric state $\psi_1(\mathbf{s}_0; \mathbf{K}^1)\propto e^{\hat{i}\Theta_1(\mathbf{s}_0)+\sum_iK_{i,i+1}^1(2s_{0,i}-1)(2s_{0,i+1}-1)/2}$, where we used $s_{0,i}$ for the variable on edge $(i,i+1)$. For the phases we assume $\Theta_{0}(\boldsymbol\sigma)= \sum_{i} \Lambda_{i}^{0}\sigma_i/2$, and similarly for $\Theta_1(\mathbf{s}_0)$. The average local energies $\langle e_i \rangle, \langle e_{i,i+1} \rangle$ can still be computed exactly for such a wave function. Note that the quantum expectation of any local observable would depend on a local subset of the parameters, thanks to the factorization property of the symmetric tree states and orthogonality of the excited states. Given the local energy functions we use the above minsum equations to find the optimal variational parameters. Figure \ref{f4} shows how such wave functions work by increasing the number of layers. The data in the figure have been obtained for $\Lambda_i=0,\pm \pi/2$ and $K_{ij}^I=0$; we did not observe significant improvement by changing these parameters, at least for the two-layer wave function. Here the relative error computed at the critical point reads $\delta e_S(1)=0.01825(t=0), 0.00542(t=1), 0.00540(t=2)$ for the $(t+1)$-layer wave functions. In the ordered phase for $h=0.9$ we obtain $\delta e_S(0.9)=0.0314(t=0), 0.0111(t=1), 0.0024(t=2)$, to be compared with the mean-field one $\delta e_{MF}(0.9)=0.0021(t=1)$. However, in the disordered phase $\delta e_S(1.1)=0.00261(t=1)$, which is much smaller than the error obtained by the mean-field wave functions for $t=2$.

\subsection{Higher dimensions}\label{S62}
Using the tree wave functions in higher dimensions is not so straightforward. Here we briefly describe a possible way of utilizing the tree states in two dimensions and leave more investigations for future studies; see also~\cite{SWVC-prl-2008}. 
Let us partition the system into pairs of spins represented by orthonormal states $|s_{\alpha}^xs_{\alpha}^y\rangle=\sum_{\sigma_{i_{\alpha}}\sigma_{j_{\alpha}}}\phi_{\alpha}(s_{\alpha}^xs_{\alpha}^y;\sigma_{i_{\alpha}}\sigma_{j_{\alpha}})|\sigma_{i_{\alpha}}\sigma_{j_{\alpha}}\rangle$ with binary variables $s_{\alpha}^{x,y} \in \{-1,+1\}$. Clearly, the mapping can be represented by any unitary transformation of the states $|\sigma_{i}\sigma_{j}\rangle$. The transformation from the physical variables $(\sigma_{i_{\alpha}},\sigma_{j_{\alpha}})$ to the auxiliary variables $(s_{\alpha}^x,s_{\alpha}^y)$ serves to reduce the entanglement between the two sets of $x$ and $y$ variables~\cite{V-prl-2007,V-prl-2008}. Then we proceed by coupling the auxiliary variables in the two sets by a symmetric state:
\begin{align}
|\Psi \rangle=\sum_{\mathbf{s}^x,\mathbf{s}^y} \frac{e^{\hat{i}\Theta(\mathbf{s}^x,\mathbf{s}^y)}}{\sqrt{2^N}}\prod_{(\alpha\beta) \in \mathcal{T}^x} \left(\frac{e^{K_{\alpha\beta}^xs_{\alpha}^xs_{\beta}^x/2}}{\sqrt{\cosh(K_{\alpha\beta}^{x,R})}} \right)
\prod_{(\alpha\beta) \in \mathcal{T}^y} \left(\frac{e^{K_{\alpha\beta}^ys_{\alpha}^ys_{\beta}^y/2}}{\sqrt{\cosh(K_{\alpha\beta}^{y,R})}} \right) |\mathbf{s}^x\mathbf{s}^y \rangle.
\end{align}
And the phase can be represented by $\Theta(\mathbf{s}^x,\mathbf{s}^y)=\sum_{\alpha}(\Lambda_{\alpha}^xs_{\alpha}^x+\Lambda_{\alpha}^ys_{\alpha}^y+\Gamma_{\alpha}^{xy}s_{\alpha}^xs_{\alpha}^y)+\cdots$ with real parameters $\Lambda_{\alpha}^{x,y}$ and $\Gamma_{\alpha}^{xy}$.     
Now it is easy to compute the average of the local energies in terms of the variational parameters. Then we use the same minsum equations given above to minimize the Hamiltonian expectation over the parameters. 
Figure \ref{f3} displays the results we obtain by such a wave function, which as expected works better than the mean-field wave function in the disordered phase.
Multilayer wave functions can be obtained by taking a superposition of the locally excited states represented by another similar wave function.

\section{Conclusion}\label{S7}
We proposed an ansatz of multilayer wave functions based on the coupling of the local excitations in the mean-field and the symmetric tree states. This allows us to compute exactly (for small number of layers) the quantum expectation of local observables, and employ an efficient message-passing algorithm to minimize the Hamiltonian expectation over the space of the variational parameters. Here we worked with the mean-field and the symmetric tree states, but the method can in principle be implemented with more complicated states after a proper characterization of the local excitations. It is the nature of these states and the local excitations that determines the minimal number of layers we need to approximate reasonably the ground state of the system. The problem is more difficult in the fermionic systems due to the global nature of the fermion sign, and it would be interesting to extend the method to deal with the non-local string interactions \cite{RZ-prb-2012}.  And similar techniques can be useful also in classical variational problems to estimate the local marginals by minimizing the free energy in the Bethe approximation for an appropriate multilayer probability distribution.

\acknowledgments
I am grateful to G. Semerjian for reading the manuscript and helpful comments. I would like  to thank J. I. Cirac and R. Orus for providing the iPEPS data in Fig.~\ref{f3}. Support from ERC Grant No. OPTINF  267915 is acknowledged.

\appendix

\section{Estimating quantum expectations for the correlated wave functions}\label{app-CW}
Consider the transverse field Ising model with Hamiltonian  
\begin{align}
H =-\sum_{(ij) \in \mathcal{E}_q} J_{ij} \sigma_i^z\sigma_j^z-\sum_{i} h_i \sigma_i^x.
\end{align} 
We group the variables according to a given partition $\{V_{\alpha}| \alpha=1,\dots,N_p\}$ and work with variables $\sigma_{\alpha} \equiv \{\sigma_i| i \in V_{\alpha}\}$. This helps to account more accurately for the short-range correlations within the groups. The partition also defines the neighborhood graph $\mathcal{E}$, where two groups $\alpha$ and $\beta$ are neighbors if there exists at least one quantum interaction $(ij) \in \mathcal{E}_q$ connecting $i \in V_{\alpha}$ and $j \in V_{\beta}$. Then a generalized mean-field state is obtained by $\psi(\boldsymbol\sigma;\mathbf{P}) \propto \prod_{\alpha} \phi_{\alpha}(\sigma_{\alpha};P_{\alpha})$, and a correlated wave function can be constructed by adding e.g. the two-body interactions
\begin{align}
\psi(\boldsymbol\sigma;\mathbf{P}) \propto \prod_{\alpha} \phi_{\alpha}(\sigma_{\alpha};P_{\alpha}) \prod_{(\alpha\beta)\in \mathcal{E}} \phi_{\alpha\beta}(\sigma_{\alpha},\sigma_{\beta};P_{\alpha\beta}).
\end{align} 
The parameters $P_{\alpha},P_{\alpha\beta}$ characterize the interactions, for instance $\phi_{\alpha\beta}= e^{\sum_{i \in V_{\alpha},j \in V_{\beta}}K_{ij}\sigma_i\sigma_j/2}$, and $\phi_{\alpha}= e^{\sum_{i \in V_{\alpha}}B_i\sigma_i/2+\sum_{i,j \in V_{\alpha}}K_{ij}\sigma_i\sigma_j/2+\cdots}$.

We also rewrite the Hamiltonian in an appropriate form $H=\sum_{\alpha}H_{\alpha}+\sum_{(\alpha\beta) \in \mathcal{E}} H_{\alpha\beta}$ 
with
\begin{align}
H_{\alpha} \equiv -\sum_{i,j \in V_{\alpha}: (ij) \in \mathcal{E}_q}J_{ij} \sigma_i^z\sigma_j^z-\sum_{i\in V_{\alpha}}h_i \sigma_i^x, \hskip1cm
H_{\alpha\beta} \equiv -\sum_{i \in V_{\alpha},j \in V_{\beta}: (ij) \in \mathcal{E}_q} J_{ij} \sigma_i^z\sigma_j^z.
\end{align}  
Then the Hamiltonian expectation reads
\begin{align}
\langle \Psi |H|\Psi \rangle = \sum_{\alpha} \langle e_{\alpha}(\sigma_{\alpha},\sigma_{\partial \alpha}) \rangle_{\mu} + \sum_{(\alpha\beta) \in \mathcal{E}} \langle e_{\alpha\beta}(\sigma_{\alpha},\sigma_{\beta}) \rangle_{\mu},
\end{align} 
where  
\begin{align}
 e_{\alpha}(\sigma_{\alpha},\sigma_{\partial \alpha})  &\equiv \sum_{\sigma_{\alpha}'} \frac{\phi_{\alpha}^*(\sigma_{\alpha}';P_{\alpha})}{\phi_{\alpha}^*(\sigma_{\alpha};P_{\alpha})}\prod_{\beta \in \partial \alpha}\frac{\phi_{\alpha\beta}^*(\sigma_{\alpha}',\sigma_{\beta};P_{\alpha\beta})}{\phi_{\alpha\beta}^*(\sigma_{\alpha},\sigma_{\beta};P_{\alpha\beta})} \langle \sigma_{\alpha}' |H_{\alpha}|\sigma_{\alpha} \rangle, \\
e_{\alpha\beta}(\sigma_{\alpha},\sigma_{\beta})  &\equiv \langle \sigma_{\alpha}\sigma_{\beta} |H_{\alpha\beta}|\sigma_{\alpha}\sigma_{\beta} \rangle,
\end{align}
and the average $\langle \cdot \rangle_{\mu}$ is computed with respect to the probability measure $\mu(\boldsymbol\sigma;\mathbf{P}) \equiv |\psi(\boldsymbol\sigma;\mathbf{P})|^2$. Here $\partial \alpha$ denotes the neighborhood set of $\alpha$ in $\mathcal{E}$, and $\sigma_{\partial \alpha}\equiv \{\sigma_{\beta}| \beta \in \partial \alpha \}$.
The Hamiltonian matrix elements are
\begin{align}
\langle \sigma_{\alpha}' |H_{\alpha}|\sigma_{\alpha} \rangle &= -\sum_{i,j \in V_{\alpha}: (ij) \in \mathcal{E}_q}J_{ij} \sigma_i\sigma_j\delta_{\sigma_{\alpha}',\sigma_{\alpha}}-\sum_{i\in V_{\alpha}}h_i \delta_{\sigma_{\alpha}',\sigma_{\alpha}^{-i}},\\
\langle \sigma_{\alpha}\sigma_{\beta} |H_{\alpha,\beta}|\sigma_{\alpha} \sigma_{\beta} \rangle &= -\sum_{i \in V_{\alpha},j \in V_{\beta}: (ij) \in \mathcal{E}_q} J_{ij} \sigma_i\sigma_j.
\end{align}
Here $\sigma_{\alpha}^{-i}$ is the same as $\sigma_{\alpha}$ except at site $i$.

We compute the quantum expectations within the Bethe approximation for the probability measure $\mu(\boldsymbol\sigma;\mathbf{P})$. To this end we need the cavity marginals $\mu_{\alpha \to \beta}(\sigma_{\alpha})$, that is the probability of having spin state $\sigma_{\alpha}$ in the absence of the interaction term $\phi_{\alpha \beta}(\sigma_{\alpha},\sigma_{\beta};P_{\alpha \beta})$. In the Bethe approximation we write this cavity marginal in terms of the neighboring cavity marginals $\{\mu_{\gamma \to \alpha}| \gamma \in \partial \alpha \setminus \beta\}$ and the local interactions:    
\begin{align}
\mu_{\alpha \to \beta}(\sigma_{\alpha}) \propto  |\phi_{\alpha}(\sigma_{\alpha};P_{\alpha})|^2\prod_{\gamma \in \partial \alpha \setminus \beta}\left( \sum_{\sigma_{\gamma}} |\phi_{\alpha \gamma}(\sigma_{\alpha},\sigma_{\gamma};P_{\alpha \gamma})|^2 \mu_{\gamma \to \alpha}(\sigma_{\gamma}) \right).
\end{align}
These recursive equation are called belief propagation (BP) equations \cite{MM-book-2009}. The equations can be solved by iteration starting from random initial marginals and updating the cavity marginals in a random sequential way according to the above equations. The solution to these equations gives the local marginals we need to compute the Hamiltonian expectation. More precisely, we have
\begin{align}
\langle e_{\alpha\beta}(\sigma_{\alpha},\sigma_{\beta}) \rangle_{\mu}= \sum_{\sigma_{\alpha},\sigma_{\beta}}\mu_{\alpha\beta}(\sigma_{\alpha},\sigma_{\beta})e_{\alpha\beta}(\sigma_{\alpha},\sigma_{\beta}),\\
\langle e_{\alpha}(\sigma_{\alpha},\sigma_{\partial \alpha}) \rangle_{\mu}= \sum_{\sigma_{\alpha},\sigma_{\partial \alpha}}\mu_{\alpha\partial \alpha}(\sigma_{\alpha},\sigma_{\partial \alpha}) e_{\alpha}(\sigma_{\alpha},\sigma_{\partial \alpha}), 
\end{align} 
where the local marginals are given by
\begin{align}
\mu_{\alpha\beta}(\sigma_{\alpha},\sigma_{\beta}) \propto  |\phi_{\alpha \beta}(\sigma_{\alpha},\sigma_{\beta};P_{\alpha \beta})|^2 \mu_{\alpha \to \beta}(\sigma_{\alpha})\mu_{\beta \to \alpha}(\sigma_{\beta}),\\
\mu_{\alpha\partial \alpha}(\sigma_{\alpha},\sigma_{\partial \alpha}) \propto |\phi_{\alpha}(\sigma_{\alpha};P_{\alpha})|^2 \prod_{\beta \in \partial \alpha} |\phi_{\alpha \beta}(\sigma_{\alpha},\sigma_{\beta};P_{\alpha \beta})|^2 \mu_{\beta \to \alpha}(\sigma_{\beta}).
\end{align}

\begin{figure}
\includegraphics[width=10cm]{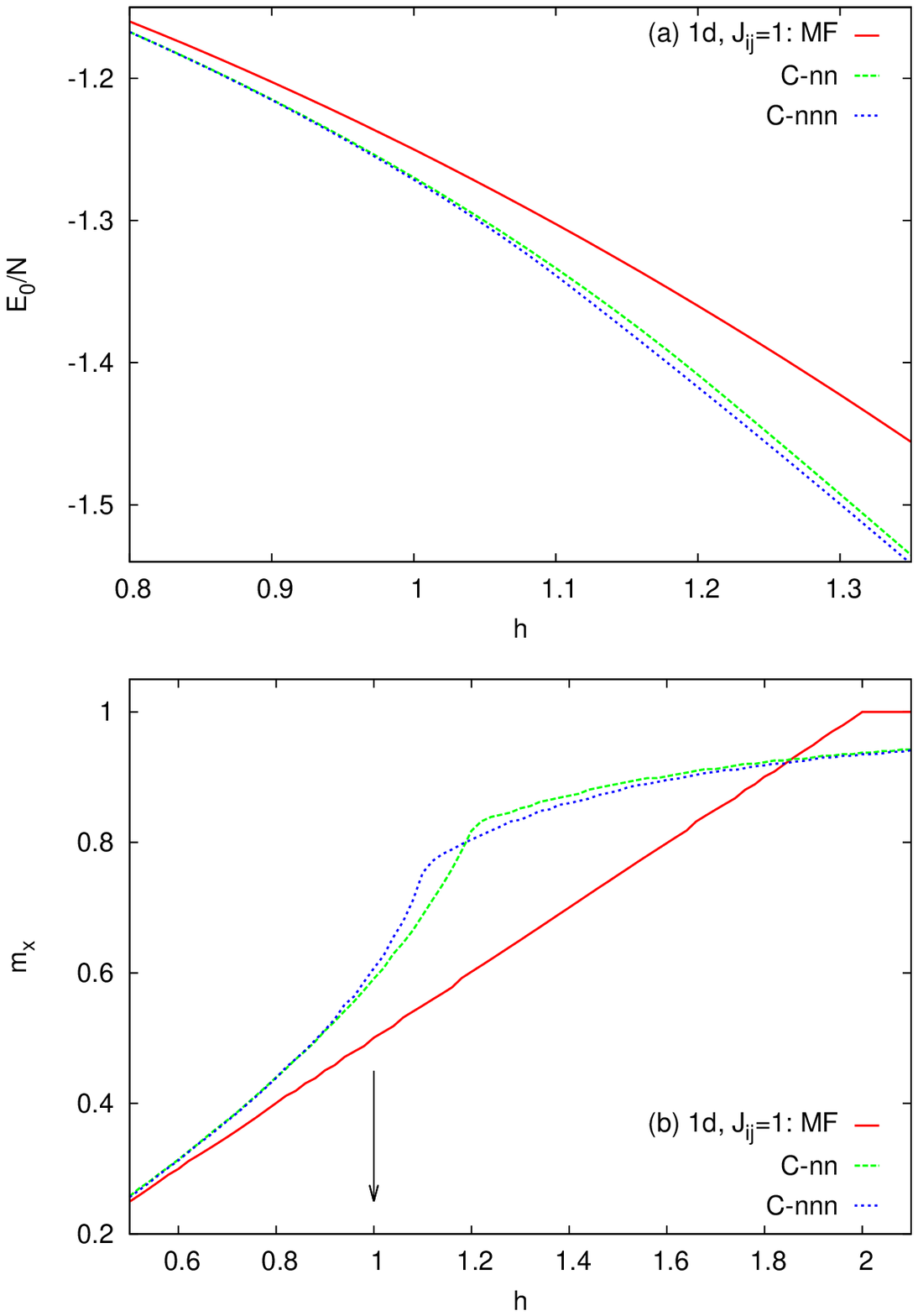}
\caption{(a) The ground-state energy density $E_0/N$, and (b) the magnetization density $m_x$ for the ferromagnetic transverse Ising model on the infinite one-dimensional lattice obtained by the translationally invariant correlated (C) wave functions. Here C-nn denotes the correlated wave functions with the nearest-neighbor interactions along the quantum interacion graph $\mathcal{E}_q$. In the C-nnn wave functins we have both the nearest-neighbor and next-nearest-neighbor interactions. The arrow shows the exact phase transition point.}\label{f5}
\end{figure}

\begin{figure}
\includegraphics[width=10cm]{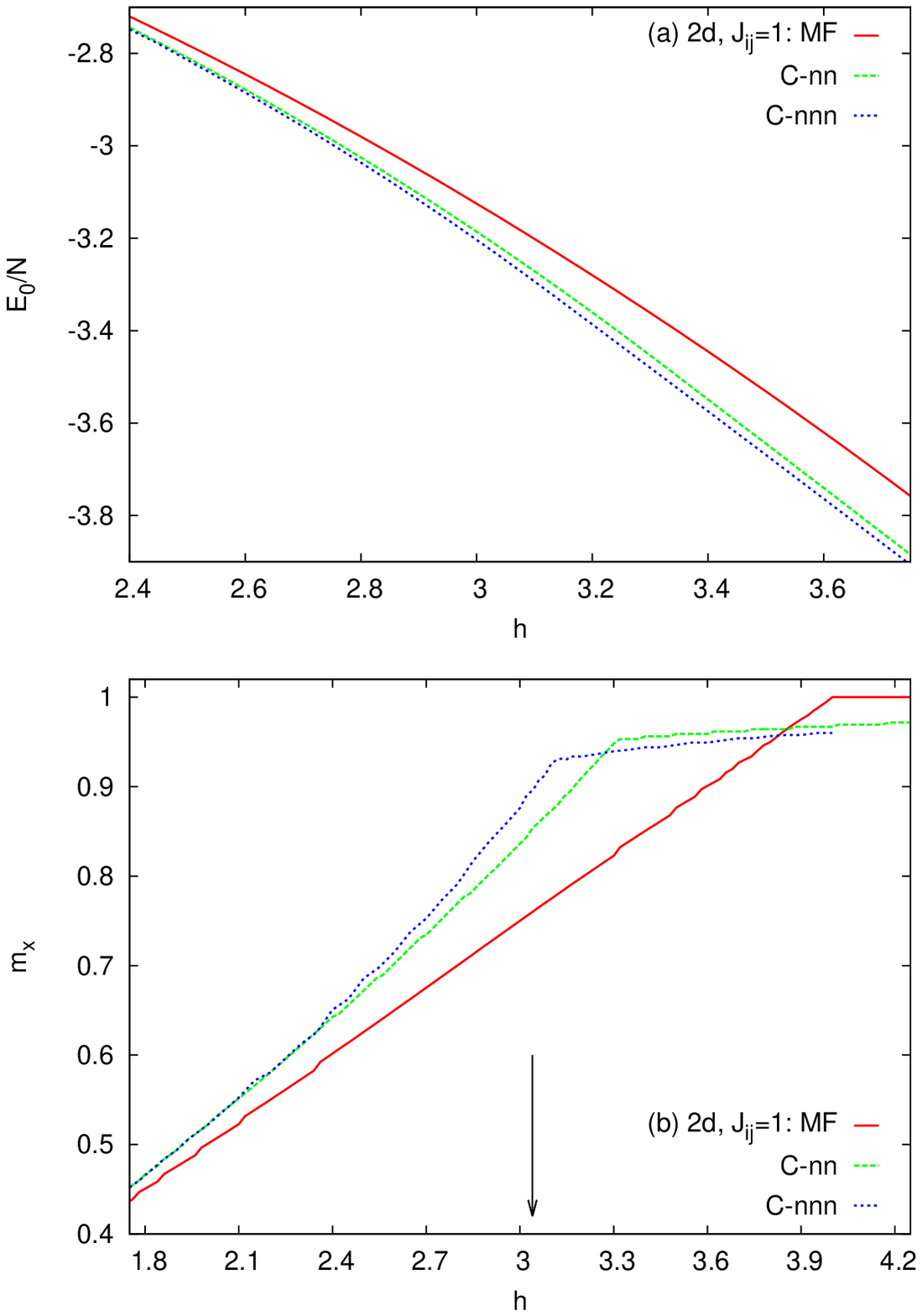}
\caption{(a) The ground-state energy density $E_0/N$, and (b) the magnetization density $m_x$ for the ferromagnetic transverse Ising model on the infinite two-dimensional square lattice obtained by the translationally invariant correlated (C) wave functions. Here C-nn denotes the correlated wave functions with the nearest-neighbor interactions along the quantum interacion graph $\mathcal{E}_q$. In the C-nnn wave functins we have both the nearest-neighbor and next-nearest-neighbor interactions. The arrow shows the expected phase transition point.}\label{f6}
\end{figure}

Now we can consider the Hamiltonian expectation as a function of the variational parameters and the cavity marginals $\mu_{\alpha \to \beta}(\sigma_{\alpha})$ satisfying the local BP equations. Then a higher-level message-passing algorithm can be used to minimize the average energy \cite{R-prb-2012}. Figures \ref{f5} and \ref{f6} show how the simple mean-field $\psi(\boldsymbol\sigma;\mathbf{P}) \propto \prod_i e^{B_i\sigma_i/2}$ and correlated wave functions $\psi(\boldsymbol\sigma;\mathbf{P}) \propto \prod_i e^{B_i\sigma_i/2}\prod_{(ij) \in \mathcal{E}} e^{K_{ij}\sigma_i\sigma_j/2}$ work in the  one- and two-dimensional transverse Ising model.

\section{Coupling the local excitations in the mean-field states}\label{app-mf}
For a mean-field state $|\Psi_0 \rangle=\sum_{\boldsymbol\sigma} \psi_0(\boldsymbol\sigma) |\boldsymbol\sigma \rangle$ with imaginary couplings,
\begin{align}
\psi_0(\boldsymbol\sigma)=e^{\hat{i}\sum_{(ij)\in \mathcal{E}_0}\Gamma_{ij}^{0}\sigma_i\sigma_j/2}\prod_i \left(\frac{e^{B_i^{0}\sigma_i/2}}{\sqrt{2\cosh(B_i^{0,R})}} \right),
\end{align}
the local energy functions are given by $e_i(\sigma_i,\sigma_{\partial_0 i})  = -  h_i e^{\hat{i}\sum_{j\in \partial_0 i}\Gamma_{ij}^{0}\sigma_i\sigma_j-(B_i^{0,R}-\hat{i}B_i^{0,I})\sigma_i}$,
and $e_{ij}(\sigma_i,\sigma_j) = -J_{ij} \sigma_i\sigma_j$.
The average local energies read
\begin{align}
\langle e_i(\sigma_i,\sigma_{\partial_0 i}) \rangle_{\mu_0} &= -  \frac{h_i}{\cosh(B_i^{0,R})}\mathrm{Re}\left\{e^{\hat{i}B_i^{0,I}}\prod_{j\in \partial_0 i}\left(\frac{\cosh(B_j^{0,R}+\hat{i}\Gamma_{ij}^{0})}{\cosh(B_j^{0,R})}\right)\right\},\\ 
\langle e_{ij}(\sigma_i,\sigma_j) \rangle_{\mu_0} &= -J_{ij} \tanh(B_i^{0,R})\tanh(B_j^{0,R}),
\end{align}
where $\langle \cdot \rangle_{\mu_0}$ means an average with respect to $\mu_0(\boldsymbol\sigma)\equiv |\psi_0(\boldsymbol\sigma)|^2$. When the transverse fields $h_i$ are nonnegative we can minimize the average energy by setting $\Gamma_{ij}^{0}=B_i^{0,I}=0$. 

We represent a locally excited state by $|\mathbf{s}_0 \rangle \equiv |s_{0,1},\dots,s_{0,N} \rangle$, with $s_{0,i} \in \{0,1\}$ to show the presence of a local excitation at site $i$.
Consider the following superposition of the locally excited states $|\Psi_1 \rangle=\sum_{\mathbf{s}_0} \psi_1(\mathbf{s}_0) |\mathbf{s}_0 \rangle$ and the associated probability measure $\mu_1(\mathbf{s}_0) \equiv |\psi_1(\mathbf{s}_0)|^2$. 
The Hamiltonian expectation with this wave function is $ \langle \Psi_1 |H| \Psi_1 \rangle =\sum_{i} \langle e_{i} \rangle_{\mu_1}+\sum_{(ij) \in \mathcal{E}_q} \langle e_{ij} \rangle_{\mu_1}$, where 
\begin{align}
e_{ij}  = -J_{ij}\sum_{\mathbf{s}_0'} \left( \frac{\psi_1^*(\mathbf{s}_0')}{\psi_1^*(\mathbf{s}_0)} \right) \langle \mathbf{s}_0' |\sigma_i^z\sigma_j^z | \mathbf{s}_0 \rangle, \hskip1cm 
 e_{i} =-h_i \sum_{\mathbf{s}_0'} \left( \frac{\psi_1^*(\mathbf{s}_0')}{\psi_1^*(\mathbf{s}_0)} \right) \langle \mathbf{s}_0' |\sigma_i^x | \mathbf{s}_0 \rangle.
\end{align}

To compute the average energies we start from $[\sigma_i^z ]^{\boldsymbol\sigma'\boldsymbol\sigma} \equiv \langle \boldsymbol\sigma' |\sigma_i^z | \boldsymbol\sigma \rangle = \sigma_i\delta_{\boldsymbol\sigma',\boldsymbol\sigma}$ and $[\sigma_i^x ]^{\boldsymbol\sigma'\boldsymbol\sigma} \equiv \langle \boldsymbol\sigma' |\sigma_i^x | \boldsymbol\sigma \rangle = \delta_{\sigma_i',-\sigma_i}\prod_{j\ne i}\delta_{\sigma_{j}',\sigma_{j}}$. 
We recall that the higher-level matrix elements are given by 
\begin{align}
[ O ]^{\mathbf{s}_0'\mathbf{s}_0} =  \sum_{\boldsymbol\sigma} |\psi_{0}(\boldsymbol\sigma,\mathbf{s}_0)|^2 
\left( \sum_{\boldsymbol\sigma'} \frac{\psi_{0}^*(\boldsymbol\sigma',\mathbf{s}_0')}{\psi_{0}^*(\boldsymbol\sigma,\mathbf{s}_0)} [ O ]^{\boldsymbol\sigma'\boldsymbol\sigma} \right),
\end{align}
where $\psi_{0}(\boldsymbol\sigma,\mathbf{s}_0)=\langle \boldsymbol\sigma |\mathbf{s}_0 \rangle$. Then one can easily obtain
\begin{align}
[\sigma_i^z ]^{\mathbf{s}_0'\mathbf{s}_0} = \prod_{j\ne i}\delta_{s_{0,j}',s_{0,j}}(1-2s_{0,i})\left(\delta_{s_{0,i}',s_{0,i}}\tanh(B_i^{0,R})-\delta_{s_{0,i}',1-s_{0,i}}\frac{\hat{i}}{\cosh(B_i^{0,R})} \right).
\end{align} 
And by the mean-field character of the wave function, we have $[\sigma_i^z\sigma_j^z ]^{\mathbf{s}_0'\mathbf{s}_0}=[\sigma_i^z ]^{\mathbf{s}_0'\mathbf{s}_0}[\sigma_j^z ]^{\mathbf{s}_0'\mathbf{s}_0}$. For the matrix elements of $\sigma_i^x$ we find
\begin{multline}
[\sigma_i^x ]^{\mathbf{s}_0'\mathbf{s}_0}  = \prod_{k\notin \{i,\partial_0 i\}}\delta_{s_{0,k}',s_{0,k}}\times \frac{1}{2\cosh(B_i^{0,R})}
 \times \\ \left(e^{f(s_{0,i},s_{0,i}')} \prod_{j \in \partial_0 i}\frac{\cosh(g(s_{0,j},s_{0,j}')+\hat{i}\Gamma_{ij}^0)}{\cosh(B_j^{0,R})}+ e^{-f(s_{0,i},s_{0,i}')}\prod_{j \in \partial_0 i}\frac{\cosh(g(s_{0,j},s_{0,j}')-\hat{i}\Gamma_{ij}^0)}{\cosh(B_j^{0,R})} \right),
\end{multline}
where we defined
\begin{align}
f(s_{0,i},s_{0,i}') &\equiv (s_{0,i}'-s_{0,i})B_i^{0,R}+\hat{i}B_i^{0,I}+\hat{i}(s_{0,i}'+s_{0,i})\pi/2, \\
g(s_{0,j},s_{0,j}') &\equiv (1-s_{0,j}'-s_{0,j})B_j^{0,R}+\hat{i}(s_{0,j}'-s_{0,j})\pi/2.
\end{align} 
When the imaginary couplings $\Gamma_{ij}^0$ and fields $B_i^{0,I}$ are zero we get
\begin{align}
[\sigma_i^x ]^{\mathbf{s}_0'\mathbf{s}_0}  =\prod_{j\ne i}\delta_{s_{0,j}',s_{0,j}} 
(1-2s_{0,i})\left(\delta_{s_{0,i}',s_{0,i}}\frac{1}{\cosh(B_i^{0,R})}+\delta_{s_{0,i}',1-s_{0,i}}\hat{i}\tanh(B_i^{0,R}) \right).
\end{align}

Finally, the average local energies are given by
\begin{align}
\langle \sigma_i^z \rangle  &=  \sum_{\mathbf{s}_0}\mu_1(\mathbf{s}_0) (1-2s_{0,i}) \left( \tanh(B_i^{0,R})- \frac{\hat{i}}{\cosh(B_i^{0,R})}\frac{\psi_1^*(\mathbf{s}_0^{-i})}{\psi_1^*(\mathbf{s}_0)} \right),
\end{align} 
\begin{multline}
\langle \sigma_i^z\sigma_j^z \rangle  = \sum_{\mathbf{s}_0}\mu_1(\mathbf{s}_0)   (1-2s_{0,i})(1-2s_{0,j}) \Big\{ \tanh(B_i^{0,R})\tanh(B_j^{0,R})\\- \frac{1}{\cosh(B_i^{0,R})\cosh(B_j^{0,R})}\frac{\psi_1^*(\mathbf{s}_0^{-i,-j})}{\psi_1^*(\mathbf{s}_0)} - \hat{i}\frac{\tanh(B_i^{0,R})}{\cosh(B_j^{0,R})}\frac{\psi_1^*(\mathbf{s}_0^{-j})}{\psi_1^*(\mathbf{s}_0)}- \hat{i}\frac{\tanh(B_j^{0,R})}{\cosh(B_i^{0,R})}\frac{\psi_1^*(\mathbf{s}_0^{-i})}{\psi_1^*(\mathbf{s}_0)} \Big\}. 
\end{multline} 
Here $\mathbf{s}_0^{-i}$ and $\mathbf{s}_0^{-i,-j}$ are configurations that are different from $\mathbf{s}_0$ only at site $i$ and sites $\{i,j\}$, respectively. 
And
\begin{multline}
\langle \sigma_i^x \rangle  = \sum_{\mathbf{s}_0}\mu_1(\mathbf{s}_0) \sum_{s_{0,i}',s_{0,\partial i}'} \left( \frac{\psi_1^*(\mathbf{s}_0| s_{0,i}',s_{0,\partial i}')}{\psi_1^*(\mathbf{s}_0)} \right)\frac{1}{2\cosh(B_i^{0,R})} \\ \times \left(e^{f(s_{0,i},s_{0,i}')}\prod_{j \in \partial_0 i}\frac{\cosh(g(s_{0,j},s_{0,j}')+\hat{i}\Gamma_{ij}^0)}{\cosh(B_j^{0,R})}+ e^{-f(s_{0,i},s_{0,i}')}\prod_{j \in \partial_0 i}\frac{\cosh(g(s_{0,j},s_{0,j}')-\hat{i}\Gamma_{ij}^0)}{\cosh(B_j^{0,R})} \right),  
\end{multline} 
where $(\mathbf{s}_0| s_{0,i}',s_{0,\partial i}')$ means we replace $s_{0,i},s_{0,\partial i}$ in $\mathbf{s}_0$ with $s_{0,i}',s_{0,\partial i}'$.

\section{Coupling the local excitations in the symmetric tree states}\label{app-s}
We take a symmetric tree state $|\Psi_0 \rangle=\sum_{\boldsymbol\sigma} \psi_0(\boldsymbol\sigma) |\boldsymbol\sigma \rangle$ with imaginary fields,
\begin{align}
\psi_0(\boldsymbol\sigma)=\frac{e^{\hat{i}\sum_i \Lambda_i^{0}\sigma_i/2}}{\sqrt{2^N}}\prod_{(ij) \in \mathcal{T}_0} \left(\frac{e^{K_{ij}^{0}\sigma_i\sigma_j/2}}{\sqrt{\cosh(K_{ij}^{0,R})}} \right).
\end{align}
The local energy functions are $e_i(\sigma_i,\sigma_{\partial_0 i})  = -  h_i e^{\hat{i}\Lambda_i^{0}\sigma_i-\sum_{j\in \partial_0 i}(K_{ij}^{0,R}-\hat{i}K_{ij}^{0,I})\sigma_i\sigma_j}$,
and $e_{ij}(\sigma_i,\sigma_j) = -J_{ij} \sigma_i\sigma_j$.
Thus for the average energies we obtain
\begin{align}
\langle e_i(\sigma_i,\sigma_{\partial i}) \rangle_{\mu_0} &= -h_i \cos(\Lambda_i^{0})\prod_{j\in \partial_0 i} \left( \frac{\cos(K_{ij}^{0,I})}{\cosh(K_{ij}^{0,R})} \right), \label{sx-1}\\ 
\langle e_{ij}(\sigma_i,\sigma_j) \rangle_{\mu_0} &= -J_{ij} \tanh(K_{ij}^{0,R}). 
\end{align}
When the transverse fields $h_i$ are nonnegative we can minimize the average energy by setting $\Lambda_i^{0}=K_{ij}^{0,I}=0$. 

Again, we represent a locally excited state by $|\mathbf{s}_0 \rangle$, with $s_{0,ij} \in \{0,1\}$ to show the presence of a local excitation at edge $(ij)$. We take a superposition of the locally excited states $|\Psi_1 \rangle=\sum_{\mathbf{s}_0} \psi_1(\mathbf{s}_0) |\mathbf{s}_0 \rangle$ and write the Hamiltonian expectation $ \langle \Psi_1 |H| \Psi_1 \rangle =\sum_{(ij) \in \mathcal{E}_q}  \langle e_{ij} \rangle_{\mu_1}+\sum_{i} \langle e_{i} \rangle_{\mu_1}$, where 
\begin{align}
e_{ij}  = -J_{ij} \sum_{\mathbf{s}_0'} \left( \frac{\psi_1^*(\mathbf{s}_0')}{\psi_1^*(\mathbf{s}_0)} \right) \langle \mathbf{s}_0' |\sigma_i^z\sigma_j^z | \mathbf{s}_0 \rangle, \hskip1cm
 e_{i} = -h_i \sum_{\mathbf{s}_0'} \left( \frac{\psi_1^*(\mathbf{s}_0')}{\psi_1^*(\mathbf{s}_0)} \right) \langle \mathbf{s}_0' |\sigma_i^x | \mathbf{s}_0 \rangle.
\end{align} 
To compute the average local energies we need the matrix elements:
\begin{align}
[\sigma_i^z\sigma_j^z ]^{\mathbf{s}_0',\mathbf{s}_0}= \prod_{(kl)\neq (ij)}\delta_{s_{0,kl}',s_{0,kl}}(1-2s_{0,ij})\left( \delta_{s_{0,ij}',s_{0,ij}}\tanh(K_{ij}^{0,R})-\delta_{s_{0,ij}',1-s_{0,ij}}\frac{\hat{i}}{\cosh(K_{ij}^{0,R})} \right),
\end{align}
and
\begin{multline}
[\sigma_i^x ]^{\mathbf{s}_0',\mathbf{s}_0}=\prod_{(kl)\notin \partial_0 i }\delta_{s_{0,kl}',s_{0,kl}}\times 
\cos(\Lambda_i^0) \prod_{j \in \partial_0 i} \Big\{ \delta_{s_{0,ij}',s_{0,ij}}\frac{\cos(K_{ij}^{0,I})}{\cosh(K_{ij}^{0,R})}(1-2s_{0,ij}) \\-\delta_{s_{0,ij}',1-s_{0,ij}}\left(\sin(K_{ij}^{0,I})-\hat{i}(1-2s_{0,ij})\cos(K_{ij}^{0,I})\tanh(K_{ij}^{0,R})\right) \Big\}.
\end{multline} 
Here by symmetry $\langle \sigma_i^z \rangle=0$. Using the above expressions the average values read 
\begin{multline}\label{sx-2}
\langle \sigma_i^x \rangle  =   \sum_{\mathbf{s}_0}\mu_1(\mathbf{s}_0)\sum_{s_{0,\partial i}'} \left( \frac{\psi_1^*(\mathbf{s}_0|s_{0,\partial i}')}{\psi_1^*(\mathbf{s}_0)} \right)\cos(\Lambda_i^0) \prod_{j \in \partial_0 i}\Big\{ \delta_{s_{0,ij}',s_{0,ij}}\frac{\cos(K_{ij}^{0,I})}{\cosh(K_{ij}^{0,R})}(1-2s_{0,ij}) \\-\delta_{s_{0,ij}',1-s_{0,ij}}\left(\sin(K_{ij}^{0,I})-\hat{i}(1-2s_{0,ij})\cos(K_{ij}^{0,I})\tanh(K_{ij}^{0,R})\right) \Big\},
\end{multline} 
and
\begin{align}
\langle \sigma_i^z\sigma_j^z \rangle  &= \sum_{\mathbf{s}_0}\mu_1(\mathbf{s}_0)   (1-2s_{0,ij}) \left( \tanh(K_{ij}^{0,R})- \frac{\hat{i}}{\cosh(K_{ij}^{0,R})}\frac{\psi_1^*(\mathbf{s}_0^{-ij})}{\psi_1^*(\mathbf{s}_0)} \right), 
\end{align} 
where $\mathbf{s}_0^{-ij}$ is different from $\mathbf{s}_0$ only at edge $(ij)$.

\end{document}